\documentclass[11pt,twoside]{article}
\usepackage{asp2014,hyperref}

\aspSuppressVolSlug
\resetcounters

\begin{document}

\title{High-speed Photometric Observations of ZZ Ceti White Dwarf Candidates}
\author{E.~M. Green,$^1$ M.-M. Limoges,$^2$ A. Gianninas,$^3$ P. Bergeron,$^2$ G. Fontaine,$^2$ P.~Dufour,$^2$ C.~J. O'Malley,$^1$ B. Guvenen,$^1$ L.~I. Biddle,$^1$ K. Pearson,$^1$ T.~W.~Deyoe,$^1$ C.~W. Bullivant,$^1$ J.~J. Hermes,$^4$ V. Van Grootel,$^5$ and M.~Grosjean$^5$
\affil{$^1$Steward Observatory, University of Arizona, 933 N. Cherry Avenue, Tucson, AZ 85721, USA; \email{bgreen@as.arizona.edu}}
\affil{$^2$D\'epartement de Physique, Universit\'e de Montr\'eal, C.~P. 6128, Succ. Centre-Ville, Montr\'eal, Qu\'ebec, H3C 3J7, Canada} %\email{AuthorEmail@email.edu}}
\affil{$^3$Homer L. Dodge Department of Physics \& Astronomy, University of Oklahoma, 440 W. Brooks St., Norman, OK 73019, USA} %\email{AuthorEmail@email.edu}}}
\affil{$^4$Department of Physics, University of Warwick, Coventry, CV4 7AL, UK}
\affil{$^5$Institut d'Astrophysique et de G\'eophysique de l'Universit\'e de Li\`ege, All\'ee du 6 Ao\^ut 17, B-4000 Li\`ege, Belgium}}
% This section is for ADS Processing.  There must be one line per author.
\paperauthor{E. M. Green}{bgreen@as.arizona.edu}{}{University of Arizona}{Steward Observatory}{Tucson}{Arizona}{85721}{USA}
\paperauthor{M.-M. Limoges}{limoges@astro.umontreal.ca}{}{Universit\'e de Montr\'eal}{D\'epartement de Physique}{Montr\'eal}{Qu\'ebec}{H3C 3J7}{Canada}
\paperauthor{A. Gianninas}{alexg@nhn.ou.edu}{0000-0002-8655-4308}{University of Oklahoma}{Homer L. Dodge Department of Physics \& Astronomy}{Norman}{Oklahoma}{73019}{USA}
\paperauthor{P. Bergeron}{bergeron@astro.umontreal.ca}{}{Universit\'e de Montr\'eal}{D\'epartement de Physique}{Montr\'eal}{Qu\'ebec}{H3C 3J7}{Canada}
\paperauthor{G. Fontaine}{fontaine@astro.umontreal.ca}{}{Universit\'e de Montr\'eal}{D\'epartement de Physique}{Montr\'eal}{Qu\'ebec}{H3C 3J7}{Canada}
\paperauthor{P. Dufour}{dufourpa@astro.umontreal.ca}{}{Universit\'e de Montr\'eal}{D\'epartement de Physique}{Montr\'eal}{Qu\'ebec}{H3C 3J7}{Canada}
\paperauthor{C. J. O'Malley}{}{}{University of Arizona}{Steward Observatory}{Tucson}{Arizona}{85721}{USA}
\paperauthor{B. Guvenen}{}{}{University of Arizona}{Steward Observatory}{Tucson}{Arizona}{85721}{USA}
\paperauthor{L. I. Biddle}{}{}{University of Arizona}{Steward Observatory}{Tucson}{Arizona}{85721}{USA}
\paperauthor{K. Pearson}{}{}{University of Arizona}{Steward Observatory}{Tucson}{Arizona}{85721}{USA}
\paperauthor{T. W. Deyoe}{}{}{University of Arizona}{Steward Observatory}{Tucson}{Arizona}{85721}{USA}
\paperauthor{C. W. Bullivant}{}{}{University of Arizona}{Steward Observatory}{Tucson}{Arizona}{85721}{USA}
\paperauthor{J. J. Hermes}{j.j.hermes@warwick.ac.uk}{}{University of Warwick}{Department of Physics}{Coventry}{}{CV4 7AL}{UK}
\paperauthor{V. Van Grootel}{valerie.vangrootel@ulg.ac.be}{}{Universit\'e de Li\`ege}{Institut d'Astrophysique et de G\'eophysique}{Li\`ege}{}{B-4000}{Belgium}
\paperauthor{M. Grosjean}{}{}{Universit\'e de Li\`ege}{Institut d'Astrophysique et de G\'eophysique}{Li\`ege}{}{B-4000}{Belgium}

\begin{abstract}
We present high-speed photometric observations of ZZ Ceti white dwarf
candidates drawn from the spectroscopic survey of bright DA stars from
the Villanova White Dwarf Catalog by Gianninas et al., and from the
recent spectroscopic survey of white dwarfs within 40 parsecs of the
Sun by Limoges et al. We report the discovery of six new ZZ Ceti
pulsators from these surveys, and several photometrically
constant DA white dwarfs, which we then use to refine the location 
of the ZZ Ceti instability strip.
\end{abstract}

\section{Introduction}

We have recently completed two major spectroscopic surveys of DA stars
using the spectroscopic approach where hydrogen Balmer lines are
fitted with the predictions of detailed model
atmospheres. \citet{GBR11} conducted a spectroscopic survey of over
1300 bright ($V <$ 17.5), hydrogen-rich white dwarfs based largely on
the last published version of the McCook \& Sion catalog
\citep{MS99}. The Gianninas et al. sample included 56 known ZZ Ceti
stars, 145 photometrically constant DA stars as well as several white
dwarfs whose atmospheric parameters placed them within or near the 
empirical boundaries of the ZZ Ceti instability strip \citep[see Figure
  35,][]{GBR11}. This includes the ultra-massive ZZ Ceti star,
GD~518 (WD~1659+662), discovered by \citet{H13}. More recently,
\citet{L13,L14} performed an exhaustive spectroscopic survey of the
SUPERBLINK proper motion database \citep[see][and references
  therein]{LS05} aimed at obtaining a complete sample of white dwarfs
in the solar neighborhood within 40 pc of the Sun.  Several ZZ Ceti
white dwarf candidates were also identified in this survey.

\section{Atmospheric Parameters}

The optical spectra for our ZZ Ceti candidates have been secured using
the Steward Observatory 2.3~m telescope equipped with the Boller \&
Chivens spectrograph. This provides wavelength coverage from
$\lambda$~$\approx$~3800--5200~\AA\ with a resolution of 6~\AA\ FWHM.

In Table 1, we present the effective temperature, $T_{\rm eff}$, and
surface gravity, $\log g$, for each star. These parameters were
measured using the standard spectroscopic technique developed by
\citet{BSL92}, with recent improvements presented in \citet{LBH05} and
\citet{GBR11}.  These fits were performed using our most recent grid
of model atmospheres, based on the ML2/$\alpha$~=~0.7 version of the
mixing-length theory.

\articlefigure[scale=0.6,bb=18 50 600 725]{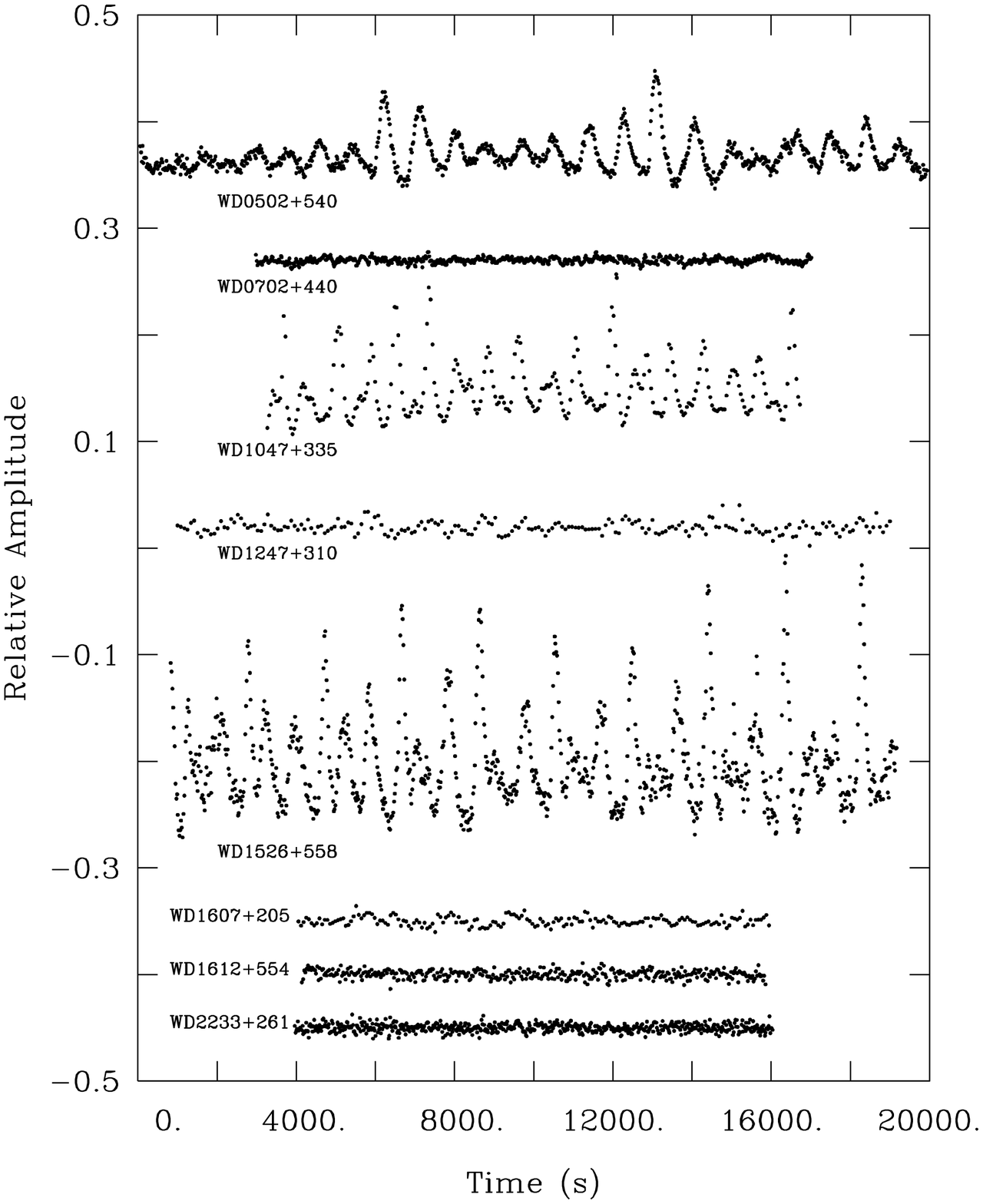}{fg:lc}
{Light curves of our ZZ Ceti white dwarf candidates.}

\begin{table}[!ht]
\caption{Observations of ZZ Ceti Star Candidates}
\smallskip
\begin{center}
{\small
\begin{tabular}{ccccccccc}  % l = left, c = centered
\tableline
\tableline
\noalign{\smallskip}
WD & $V$ or $B$ & $T_{\rm eff}$ & $\log g$ & Length & Bandpass & Period & Amplitude & 4$\sigma$ \\
   & (mag)      & (K)           &          & (hr)   & (mHz)    & (s)    & (\%)      & (\%)      \\
\noalign{\smallskip}
\tableline
\noalign{\smallskip}
0502+540 & 15.2 & 11400 & 8.24 &  9.18 & 0--5   &  873.6 & 1.27 & 0.41\\
0702+440 & 15.1 & 11000 & 8.29 & 15.05 & 0--10  & 1366.4 & 0.10 & 0.07\\
1047+335 & 17.0 & 11430 & 8.24 &  4.59 & 0--10  &  767.5 & 2.77 & 0.25\\
1247+310 & 17.2 & 12110 & 8.43 &  9.76 & 0--4.5 &  364.6 & 0.20 & 0.20\\
1526+558 & 17.0 & 11020 & 7.89 &  5.14 & 0--10  &  648.9 & 3.68 & 0.26\\
1607+205 & 17.4 & 11280 & 7.94 &  9.35 & 0--7.5 & 1928.5 & 0.18 & 0.13\\
1612+554 & 16.5 & 12100 & 8.33 &  3.26 & 0--10  &  NOV & \ldots & 0.14\\
2233+261 & 15.3 & 12020 & 8.16 &  9.82 & 0--10  &  NOV & \ldots & 0.07\\
\\
1659+622 & 16.0 & 13050 & 8.06 &  6.66 & 0--10  &  NOV & \ldots & 0.12\\
\noalign{\smallskip}
\tableline
\tableline
\end{tabular}
}
\end{center}
\end{table}

\section{High-Speed Photometry and Fourier Analysis}

We obtained high-speed photometric measurements of our ZZ Ceti white 
dwarf candidates using Steward Observatory's 1.55~m Kuiper telescope 
on Mt. Bigelow and the Mont4K CCD camera\footnote{Please consult the 
following Web site:
  \url{http://james.as.arizona.edu/~psmith/61inch/instruments.html} for more
  details, if interested.} through a Schott-8612 ``white light'' filter, 
with the exception of WD~1526+558, which was observed at the Kitt Peak 
2.1~m telescope. The observed light curves of our ZZ Ceti white dwarf 
candidates (except WD~1659+622), are shown in Figure~\ref{fg:lc}. The 
total time spent on each target is given in Table 1.

We then proceeded to compute the Fourier transforms of our light-curves 
in order to perform preliminary frequency extraction for the variable 
stars, and obtain limits on the non-variability for the photometrically 
constant stars. The results of this analysis are presented in Table 1 
where we list the observed bandpass along with the dominant period and 
its amplitude, expressed as a percentage of the mean brightness of 
the star. We also list our detection threshold which corresponds to 
four times the mean noise level within the observed bandpass ($4\sigma$).

\articlefigure[scale=0.55,angle=-90,bb=92 46 546 784]{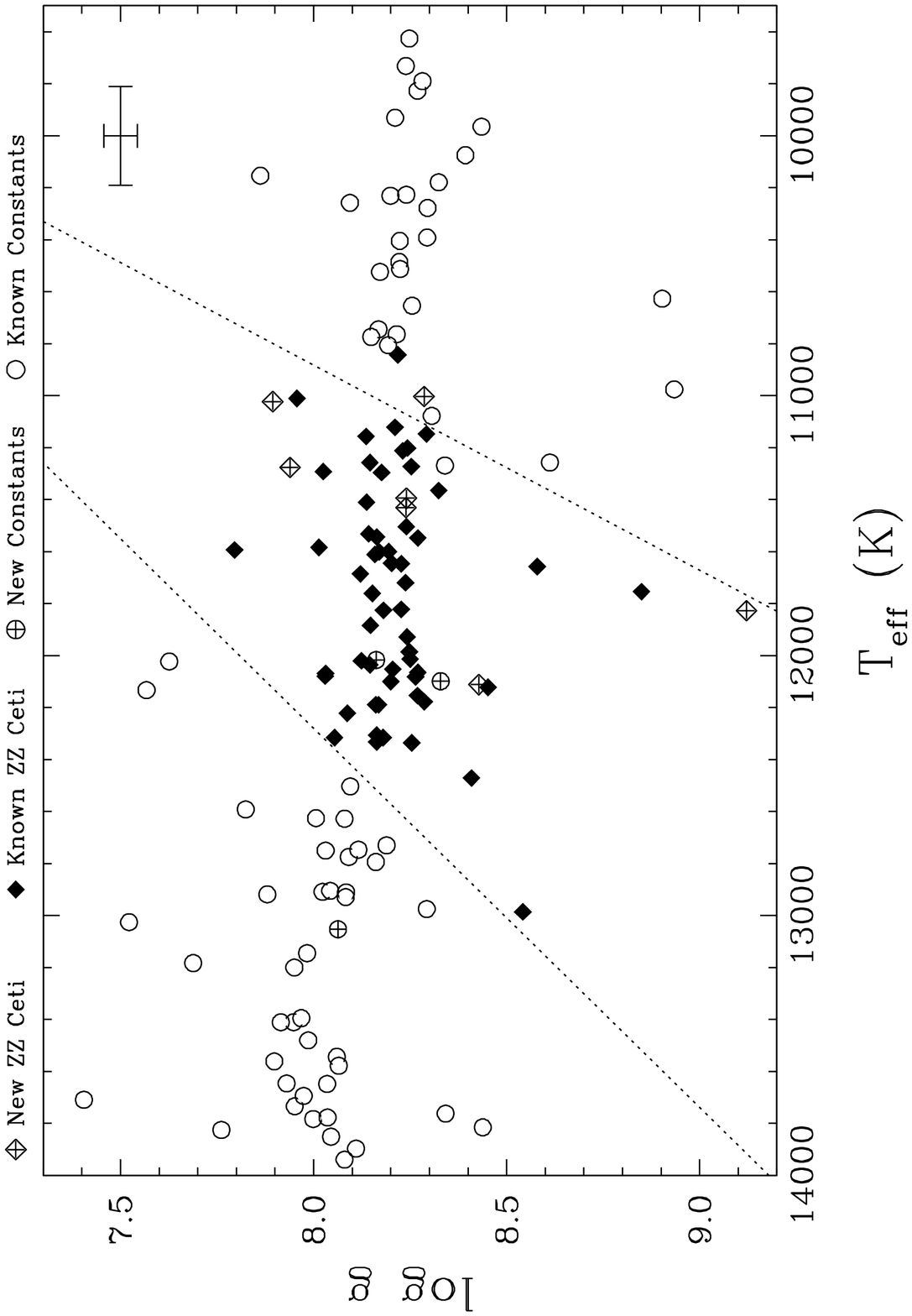}{fg:zz}
{Effective temperature vs.\ surface gravity distribution for DA white
  dwarfs with high-speed photometric measurements.}

\section{Updated ZZ Ceti Instability Strip}

Our updated $T_{\rm eff}$ vs. $\log g$ distribution for DA white
dwarfs with high-speed photometric measurements is plotted in 
Figure~\ref{fg:zz}. The new/known ZZ Ceti stars and new/known constant 
stars are indicated by different symbols (see the legend at the top of 
the figure).  The dotted lines represent the approximate empirical
boundaries of the instability strip based on our new results. The
error bars represent the average uncertainties of the spectroscopic
method in the region of the ZZ Ceti instability strip. The high-log g
ZZ Ceti star at the bottom of the plot, GD~518, has been reported by
\citet{H13}.

There are two non-variable stars located in the middle of the
instability strip. The first one is WD~2233+261, and good photometric
limits were obtained (see Table~1). However, only a single spectrum was
secured for this object, which thus deserves to be further
investigated.

The other offending star is HS~1612+5528 (WD~1612+554), one of the two
NOV white dwarfs lying in the heart of the instability
strip. Independently observed several times in spectroscopy and
high-speed photometry \citep[see][]{GBR11}, HS~1612+5528 could
represent the first ZZ Ceti star whose pulsations are hidden from us
due to geometric considerations.  Alternatively, it could be a double
degenerate binary, with individual components located outside the
instability strip. This object also merits further study.

\acknowledgements We thank the director and staff of Steward Observatory 
and Kitt Peak National Observatory for the use of their facilities. This 
work is funded in part by the NSERC Canada and by the Fund FRQ-NT 
(Qu\'ebec).

%\bibliography{editor}  % For BibTex

% For non-BibTex:

\end{document}